# Spin-to-charge conversion modulated by chiral molecules

*Peng-Yi Liu,*[1] *Tian-Yi Zhang,*[1] *Ai-Min Guo,*[2] *Yossi Paltiel,*[3,4] *Qing-Feng Sun*[1,5*]

[1]International Center for Quantum Materials, School of Physics, Peking University, Beijing, 100871, China.

[2]Hunan Key Laboratory for Super-microstructure and Ultrafast Process, School of Physics, Central South University, Changsha 410083, China.

[3]Institute of Applied Physics, The Hebrew University of Jerusalem, Jerusalem 9190401, Israel.

[4]Center for Nanoscience and Nanotechnology, The Hebrew University of Jerusalem, Jerusalem 9190401, Israel.

[5]Hefei National Laboratory, Hefei 230088, China.

*Corresponding author: sunqf@pku.edu.cn

ABSTRACT. Molecular chirality and electron spin are intricately intertwined via the fascinating phenomenon of chiral-induced spin selectivity (CISS), which has garnered considerable attention due to its extensive potential applications. A recent experiment has revealed that chiral molecules self-assembled on the gold surface can modulate the inverse spin Hall effect, providing an alternative platform for studying the interplay between chirality and spin transport. Our study uncovers that this modulation stems from the CISS effect, which enhances spin currents of one spin orientation while suppressing those of the opposite spin orientation. We provide numerical results that are highly consistent with the experimental phenomena and further investigate the influence of various factors on this modulation. This work offers a theoretical explanation of previously unexplained experimental findings, and the underlying physical mechanism broadens current perspectives on understanding and applying CISS.

TEXT.

Molecular chirality, a property attracting increasing attention, is pivotal in many chemical reactions and biological processes. When spin-unpolarized electrons transmit through chiral materials, they become highly spin-polarized, which has been substantiated by numerous experiments since 1999 and is known as chiral-induced spin selectivity (CISS).[1–3] CISS embodies the profound connection between molecular chirality and electron spin, and research in this area has gained popularity in recent years.[4–9] It is widely accepted that spin-orbit coupling (SOC) plays an essential role in CISS.[10–15] Some research attributes the spin selectivity to the chiral molecule/substrate interface.[16,17] Additionally, other studies suggest that the electron-electron interaction,[18] electron-phonon interactions,[19] and spin-dependent scattering[20] could also be responsible for CISS. Although a universal theory that fully comprehends the CISS effect remains unclear, the huge research potential of CISS has inspired many experimental[21–24] and application[25–27] investigations.

A recent experiment has established an extra framework for investigating CISS,[28] promising to unlock the applications of chiral molecules in high-dimensional systems and hybrid systems. In this experiment, Moharana *et al.* grew a thin layer of gold (Au) on a ferromagnetic substrate and employed ferromagnetic resonance (FMR) as a spin pump. This setup injects electrons with spins parallel to the external magnetic field into the gold and ejects those with spins antiparallel to the external magnetic field out of the gold, creating a spin current that conveys spin angular momentum.[29] Due to the strong spin-orbit coupling in gold, electrons with opposite spin orientations undergo deflection in opposite directions. Consequently, a charge current passing through gold can cause a transverse spin current, which is the spin Hall effect (SHE).[30] Conversely, in that experiment, FMR pushes a spin current passing through gold causing a transverse charge current, leading to a Hall voltage, which is the inverse spin Hall effect (ISHE).[29]

Results from the recent experiment[28] show that for the bare gold layer, the sign of the Hall voltage reverses upon reversing the magnetic field, while the absolute values remain equal. In contrast, after self-assembling chiral molecules (specifically, α-helix polyalanines) on the surface of the gold film, the sign of the Hall voltage reverses upon reversing the magnetic field, but the absolute values are no longer equal. In other words, there is an asymmetry in the spin-to-charge conversion process induced by the self-assembled chiral molecules. Moreover, the experiment demonstrates that the asymmetry induced by L-rotation α-helix (L-α-helix) and D-rotation α-helix (D-α-helix) polyalanines is the opposite, whereas racemic mixtures do not induce any asymmetry. In contrast to the previous chiral-related electric Hall effects, which need charge current,[27,31] this experiment is more interesting because the ISHE is caused by pure spin current. Despite these fascinating findings and their strong correlation with CISS, the underlying theoretical mechanism remains unclear, inscrutable, and requires further elucidation.

In this Letter, we focus on an FMR-SHE-chiral molecules system (see Figure 1a), identical to the experiment,[28] and investigate the modulation of chiral molecules on the spin-to-charge conversion. We demonstrate that a spin current is driven from the ferromagnetic substrate into the gold by FMR. Due to the CISS effect, chiral molecules can enhance the spin current with spins aligned in one direction while hindering those aligned in the opposite direction, which is the origin of the asymmetric spin-to-charge conversion observed in the experiment. Furthermore, we found that reversing the chirality of the molecules also reverses their modulation effect and studied the dependence of the modulation strength on system parameters. Notably, we fully explain this experiment using the CISS effect of chiral molecules, and the numerical results are highly consistent with the experimental observations. The ISHE experiment and our work together demonstrate a feasible approach to regulating spin currents at room temperature.

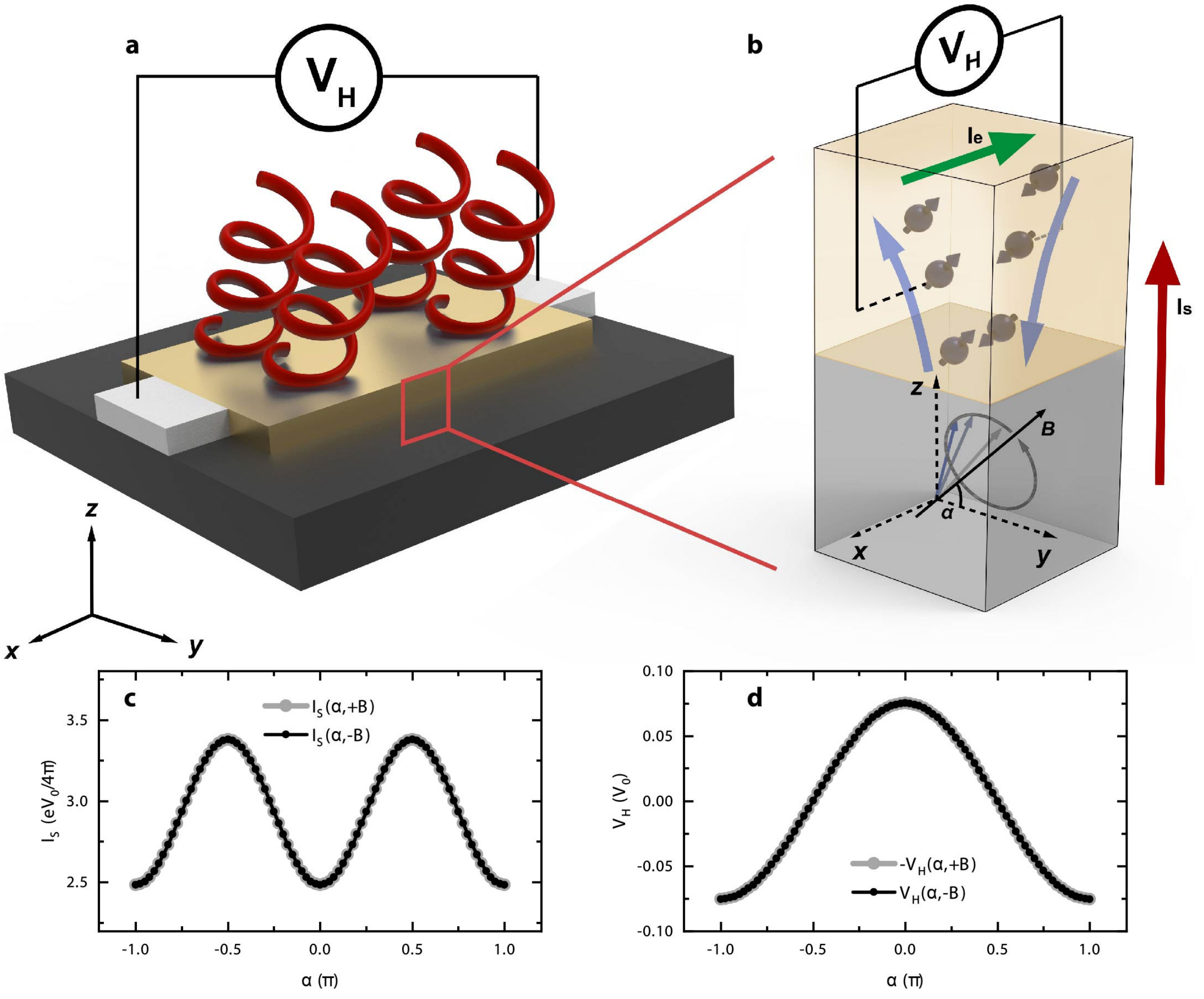

**Figure 1**. The schematic diagrams of the model and ISHE. (a) The schematic diagrams of the ISHE modulated by chiral molecules. The ferromagnetic substrate, a thin layer of gold, L-α-helix molecules, and the electrodes measuring the Hall voltage are shown in black, gold, red, and white, respectively. (b) is the schematic diagram of the ISHE, which is an enlarged diagram of the microscopic details of (a). FMR in the ferromagnetic substrate (shown in grey) causes the precession of the magnetization and injects the spin current $I_s$ (shown in the red arrow) with the spin angular momentum parallel to $\boldsymbol{B}$ into the gold, where $\boldsymbol{B}$ is the external magnetic field controlling the spin direction. In the experiment and our discussion, $\boldsymbol{B}$ is in the $yz$-plane and forms an angle of $\alpha$ with the $y$-axis. Grey dots with opposite arrows represent electrons with spins parallel and antiparallel to $\boldsymbol{B}$ in the gold (shown in gold) driven by FMR and deflecting in opposite directions, resulting a Hall charge current $I_e$ (the green arrow) and Hall voltage. (c) The amplitude of spin current $I_s(\alpha, +B)$ (grey curve) injected into the gold and its counterpart with a reversed magnetic field $I_s(\alpha, -B)$ (black curve). (d) The minus Hall voltage $-V_H(\alpha, +B)$ (grey curve) and the Hall voltage with a reversed magnetic field $V_H(\alpha, -B)$ (black curve). (c) and (d) are calculated in the absence of chiral molecules. The parameters used here are provided in Section S1 of the Supporting Information.

To theoretically investigate the experiment and elucidate the underlying mechanisms, we construct a tight-binding model capturing the key components of the experimental device: a ferromagnetic substrate performing FMR, a metallic (gold) layer exhibiting ISHE, two voltage probes measuring the Hall voltage, and a self-assembled monolayer of chiral molecules. Each part of the system and their coupling are modeled explicitly, enabling standard spin transport simulations. Full details of the model and numerical implementation are provided in Sections S1 and S2 of the Supporting Information.

First, we simulate the conventional ISHE in the absence of chiral molecules. Consider the same setup as the experiment,[28] which consists of a ferromagnetic substrate coated with a thin layer of gold (Figure 1a but without chiral molecules). In the experiment, a microwave was applied to the ferromagnetic substrate to excite the precession of magnetization direction around the external magnetic field $\boldsymbol{B}$. During this process, spins accumulate at the ferromagnetic/gold interface. Electrons with spins parallel to $\boldsymbol{B}$ experience an increase in their electrochemical potential by $eV_0$, while those with antiparallel spins undergo a decrease by $eV_0$, resulting in a Fermi energy splitting $\Delta\mu = 2eV_0 \approx \hbar\omega$.[32] Here, e and $\hbar$ represent the elementary charge and the reduced Planck constant, respectively. $V_0$ is the spin bias[33] that can drive a spin current. Given that the frequency of FMR $\omega \approx 2\pi \times 10^{10}$Hz,[28] $V_0$ is on the order of ten microvolts.

In the model (with details in Section S1 of Supporting Information), the ferromagnetic substrate and the magnetic field cause a Zeeman splitting along the direction of $\boldsymbol{B}$. More importantly, we set the ferromagnetic substrate to have the spin bias $V_0$, which drives a spin current $I_s$ flowing from the ferromagnetic substrate into the gold. The spin current is a tensor and comprises both the direction of spin angular momentum and the direction of current.[29,34] In this device, the direction of current is from the ferromagnetic substrate towards the gold ($+z$ direction), while the direction of the spin angular momentum is controlled by $\boldsymbol{B}$. In other words, electrons with spin parallel to $\boldsymbol{B}$ are pushed into the gold layer, while those with spin antiparallel to $\boldsymbol{B}$ are pulled out from the gold layer, as shown in Figure 1b.

When the spin current passes through a gold layer, as shown in Figure 1b, electrons with opposite spins deflect in opposite directions, leading to a Hall charge current ($I_e$ in Figure 1b) due to ISHE. This results in a measurable Hall voltage at the electrodes in the $x$-direction (perpendicular to both the direction of the spin current and the direction of the spin angular momentum), depicted in Figures 1a and 1b. We adopt the Bernevig-Hughes-Zhang model[35] to simulate the ISHE in the gold layer (see Section S1 of the Supporting Information and Figure S1). Utilizing the non-equilibrium Green's function method, we calculate not only the Hall voltage $V_H$ (as the experiment measured) but also the magnitude of the spin current $I_s$ which is challenging to measure experimentally (see Section S1 of the Supporting Information for details).

Figure 1c illustrates the spin current under different directions of the magnetic field without the coupling of the chiral molecules $I_s(\alpha, +B)$, where $+B$ signifies that the magnetic field used here is positive, corresponding to the reversed magnetic field $-B$ discussed later. As in the experiment, the magnetic field $\boldsymbol{B}$ is set in the $yz$-plane and $\alpha$ represents the angle between the direction of $\boldsymbol{B}$ and the $y$-axis. For example, when $\alpha = 0$ ($\pi/2$), FMR pumps spin current with the spin angular momentum point to the $y$- ($z$-) direction. For varying $\alpha$, $I_s$ fluctuates but remains positive, as the spin current is driven into the gold layer by the spin bias regardless of the spin directions. Figure 1d shows the minus Hall voltages $-V_H(\alpha, +B)$, as the experiment measured.[28] The charge current induced by ISHE is proportional to a cross product[29,30] between the spin current (point to $+z$) and the spin angular momentum (parallel to $\boldsymbol{B}$). Therefore $V_H(\alpha, +B)$ has an angle-dependent factor $\propto \hat{\boldsymbol{z}} \times \widehat{\boldsymbol{B}} = \sin(\pi/2 - \alpha)$, where $\hat{\boldsymbol{z}}$ ($\widehat{\boldsymbol{B}}$) represents the unit vector point to $+z$ (parallel to $\boldsymbol{B}$). Therefore, $-V_H(\alpha, +B)$ reaches its maximum (minimum) at $\alpha = 0$ ($\pm\pi$), vanishes and changes sign at $\alpha = \pm\pi/2$, due to the changing direction of the spin angular momentum, which is the same as the experimental results.[28]

In addition, we also present the spin current and Hall voltage when $\boldsymbol{B}$ is reversed (denoted by $-B$). The spin current is still positive for all $\alpha$ [$I_s(\alpha, -B)$ shown in Figure 1c], but the Hall voltage changes sign since the opposite direction of the spin angular momentum [$V_H(\alpha, -B)$ shown in Figure 1d]. More importantly, the relationships $I_s(\alpha, +B) = I_s(\alpha, -B)$ and $-V_H(\alpha, +B) = V_H(\alpha, -B)$ hold exactly protected by a symmetry (see Section S2 of the Supporting Information for the details). That is to say, when $\boldsymbol{B}$ is reversed, $I_s$ remains unchanged, $V_H$ is reversed and its absolute value remains unchanged, while without the coupling of chiral molecules, which is well consistent with the recent experiment.[28]

To further investigate the role of molecular chirality, we now incorporate chiral molecules into the model. As in the experiment, we consider a layer of chiral molecules self-assembled on the gold surface, as illustrated in Figures 1a and S1. Given that the polyalanines employed in the experiment possess a single-helical structure, we correspondingly use a model of $\alpha$-helical molecules for simulation, which has chirality and has proven successful in capturing the essential physics of CISS.[14,36–38] It demonstrates that spin-unpolarized charge currents become spin-polarized after passing through chiral molecules;[14] it reproduces the recent CISS experiments in electron donor–acceptor systems[6] and reveals the underlying spin-selective dynamics;[37] and it also successfully explains many magnetic phenomena induced by CISS.[38] These prior results validate the model as a powerful framework for simulating CISS-related phenomena.

The detailed Hamiltonian is given by $H_M = H_{mol} + H_{SOC}$:

$$\begin{aligned} H_{mol} &= \sum_{n=1}^{N_p} \left[ \sum_{l=1}^{L_p} d_{n,l}^{+} \left( E_p s_0 - M s_\alpha \right) d_{n,l} + \sum_{l=1}^{L_p-1} \sum_{j=1}^{L_p-l} t_j d_{n,l}^{+} s_0 d_{n,l+j} + h.c. \right], \\ H_{SOC} &= \sum_{n=1}^{N_p} \sum_{l=1}^{L_p-1} \sum_{j=1}^{L_p-l} 2 i t_{sj} \cos \varphi_{\overline{lj}} \, d_{n,l}^{+} s_{lj} d_{n,l+j} + h.c., \end{aligned} \tag{1}$$

where $d_{n,l}^{+} = [d_{n,l,\uparrow}^{+}, d_{n,l,\downarrow}^{+}]$ is the creation of electrons at site $l$ of the $n$-th molecule, with $\uparrow/\downarrow$ representing the spin degree of freedom. Each molecule has $L_p$ sites (i.e., the molecule length is $L_p$), and there are $N_p$ molecules coupled to the surface of the gold. $M$ is the Zeeman term. $E_p$ is the on-site energy of molecules. $t_j$ and $t_{sj}$ are the hopping and SOC between site $l$ and site $l + j$, respectively.$\varphi_{\overline{lj}} = j\Delta\varphi/2$, where $\Delta\varphi$ is the twist angle between the two nearest sites, determining the chirality ($\Delta\varphi = 5\pi/9, -5\pi/9, 0$ for L-α-helix, D-α-helix, and achiral molecules, respectively).

$s_{lj}$ is a $2 \times 2$ matrix related to the geometry of the molecule.[14] The model and parameters details can be found in Section S1 of the Supplementary Information.

Similar to the case without molecules, we calculate the variations of the spin current $I_s(\alpha, +B)$ passing through the ferromagnetic/gold interface and the minus Hall voltage $-V_H(\alpha, +B)$ with $\alpha$ and compare them with those under the reversed magnetic fields. The numerical results are shown in Figure 2.

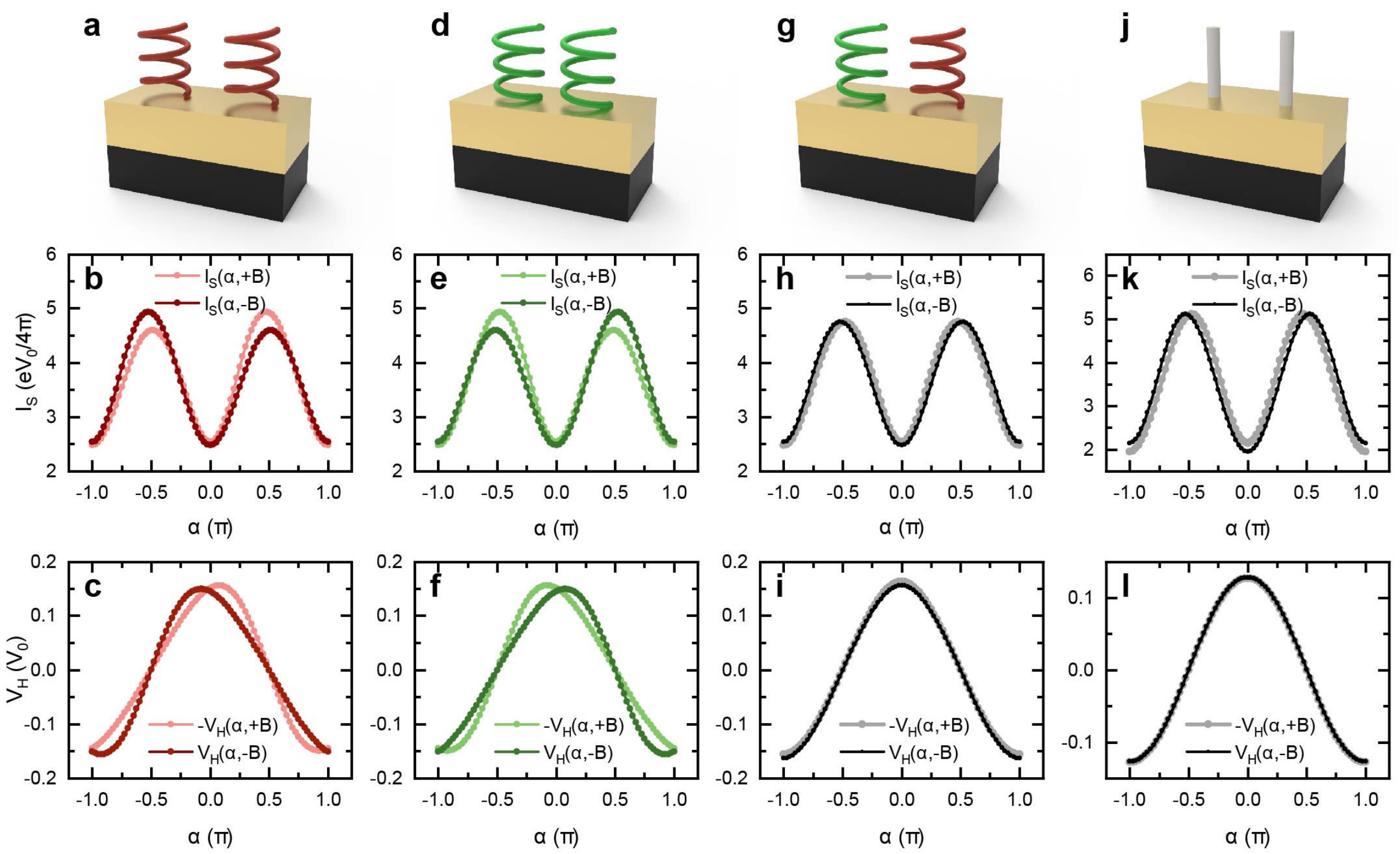

**Figure 2** The influence of chiral molecules on the ISHE. (a, d, g, j) The schematic diagrams, (b, e, h, k) spin currents $I_s$, and (c, f, i, l) Hall voltages $V_H$ of four different situations. In (a-c) and (d-f), we perform the calculations with the self-assembling of the L-α-helix and D-α-helix molecules, respectively. In (g-i) and (j-l), the racemic mixture and achiral molecules are respectively used as two control groups. Data with positive magnetic field [$I_s(\alpha, +B)$ and $-V_H(\alpha, +B)$] are shown in light red, green, and grey curves for L-α-helix molecules, D-α-helix molecules, and control groups, respectively. Data with reversed magnetic field

$[I_s(\alpha,-B)$ and $V_H(\alpha,-B)]$ are shown in dark colors, correspondingly. Parameters used here and the details of control groups (g-l) are provided in Supporting Information.

First, the magnitude of $I_s$ and $V_H$ in Figure 2 are larger than those in Figure 1 due to the flow of spin current into molecules. More importantly, as shown in Figures 2b and 2c, an asymmetry appears when L-α-helix molecules are coupled to the gold surface (see Figure 2a). Specifically, the relations $I_s(\alpha,+B) = I_s(\alpha,-B)$ and $-V_H(\alpha,+B) = V_H(\alpha,-B)$ no longer hold, which is consistent with the experimental results [see Figure S2 and Ref. [28]]. Due to the CISS effect, the L-α-helix molecules can facilitate the spin current with the spin angular momentum point toward the $+z$-direction while hindering that with the spin angular momentum point toward the $-z$-direction. Consequently, at $\alpha = \pi/2$ ($\boldsymbol{B}$ aligns with $+z$), spin currents with $+B$ propagate through molecules more readily than those with $-B$, resulting in $I_s(\pi/2,+B) > I_s(\pi/2,-B)$ (see Figure 2b). A larger spin current can cause a larger Hall effect, so around $\alpha = \pi/2$ $(0 < \alpha < \pi)$, the absolute values of Hall voltages with $+B$ are larger than those with $-B$, i.e., $|V_H(\alpha,+B)| > |V_H(\alpha,-B)|$ as shown in Figure 2c. Similarly, at around $\alpha = -\pi/2$ $(-\pi < \alpha < 0)$ the asymmetry reverses due to the opposite direction of $\boldsymbol{B}$, resulting in $I_s(\alpha,+B) < I_s(\alpha,-B)$ and $|V_H(\alpha,+B)| < |V_H(\alpha,-B)|$. When $\alpha = 0$ and $\pm\pi$, $\boldsymbol{B}$ is perpendicular to the helix axis of chiral molecules, and CISS cannot select the spin currents with $+B$ or $-B$, leading to the disappearance of the asymmetry. When $\alpha = \pm\pi/2$, $\boldsymbol{B}$ is in $\pm z$ direction and the asymmetry is the largest for $I_s$ due to the CISS effect of the chiral molecules, but the asymmetry for $V_H$ disappears with the vanished Hall voltage $V_H(\alpha,+B) = V_H(\alpha,-B) = 0$, since $\boldsymbol{B}$ parallel to $\pm z$ direction cannot cause the Hall voltage as we discussed above. Therefore, the most notable asymmetry for $V_H$ appears when $\alpha \approx \pm\pi/4, \pm 3\pi/4$. All these details in Figures 2b and 2c are in good agreement with the experiment (see Figure S2), and we conclude that the asymmetries in Ref. [28] arise from the CISS effect of chiral molecules.

In Figures 2d-2f, we consider D-α-helix molecules as a comparison for Figures 2a-2c. One of the characteristics of CISS is that molecules with opposite chirality exhibit opposite spin selectivity. So, D-α-helix molecules modulate the spin currents and Hall voltages oppositely to L-α-helix molecules, leading to the emergence of an opposite asymmetry. For example, when $0 < \alpha < \pi$, $I_s(\alpha, +B) < I_s(\alpha, -B)$, resulting in $|V_H(\alpha, +B)| < |V_H(\alpha, -B)|$. These results also align well with experimental observations (as shown in Figure S2), and the opposite modulation effects of different chiralities support the role of CISS in this process.

To further ensure that the modulation of Hall voltages arises from molecular chirality, we consider two control groups as in the experiment: the racemic mixture of chiral molecules [shown in Figures 2g-2i] and achiral molecules [shown in Figures 2j-2l]. In the experiment, the racemic mixture consists of equal fractions of both optical rotations.[28] So, we assemble mixed L-α-helix and D-α-helix molecules in equal proportions on the gold surface to simulate the racemic mixture. Specifically, we employ 1000 sets of random configurations to perform the numerical calculations and use the average results (see Figure S3 and Section S3 of the Supporting Information for the details). In each configuration, each molecule has a 50% probability of being L-α-helix and a 50% probability of being D-α-helix. The average results and the results with the error bar are shown in Figures 2h, 2i and S3. The data under $\pm B$ exhibits only little discrepancies and fails to show chiral-induced asymmetry in ISHE, which is consistent with the experiment.[28] As another control group, we use simple one-dimensional-chain molecules to simulate achiral molecules, assembled on the gold surface in the same way. The results also demonstrate no chirality-induced asymmetry (Figures 2k and 2l). Although both control groups successfully confirm that the asymmetry described above originates from chirality, the expected relations $I_s(\alpha,+B) = I_s(\alpha,-B)$ and $-V_H(\alpha,+B) = V_H(\alpha,-B)$ have slight deviations. We attribute these deviations to the mismatch between the SHE model (containing pseudospin) and the molecular model (without pseudospin). Because this can be overcome by a refined matching of the SHE and molecular models, as shown in Figure S4c. In real systems, the atomic orbitals of the molecules and substrate differ greatly, and similar mismatches inevitably exist, resulting in finite deviations in the control experiment as well.[28] More detailed discussions of control groups are provided in Figure S3, Figure S4, and Section S3 of the Supporting Information.

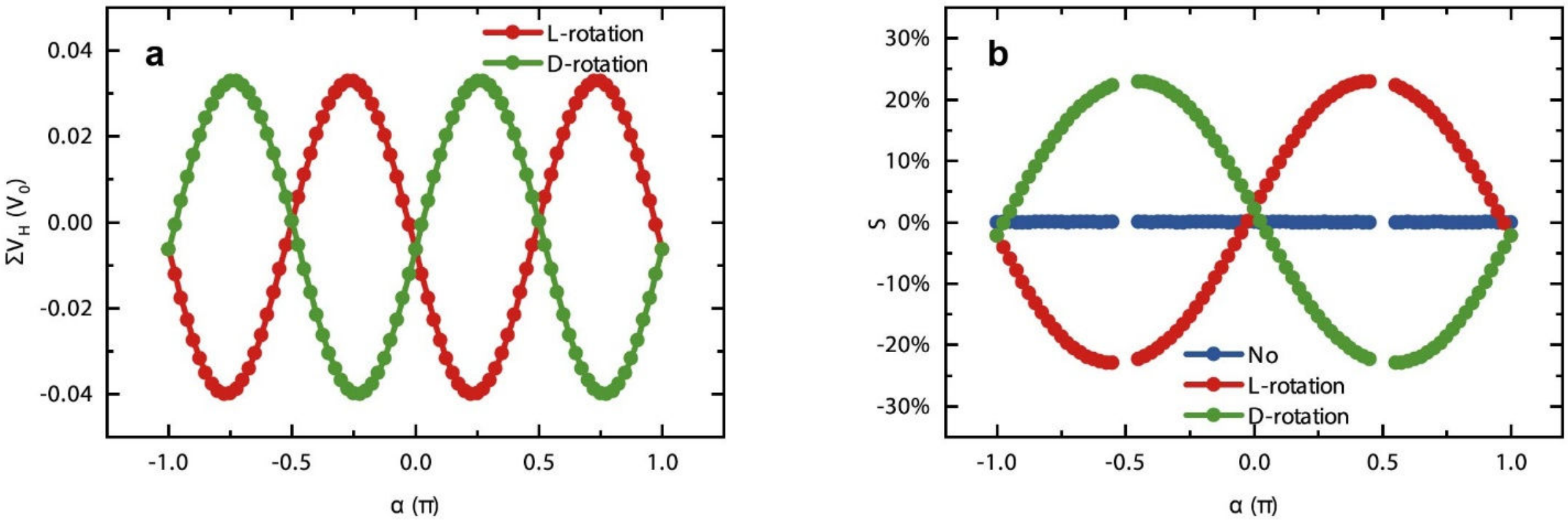

**Figure 3** The modulation strength of chiral molecules on the ISHE. (a) The summation of Hall voltages under opposite magnetic fields $\Sigma V_H(\alpha)$ with L-α-helix and D-α-helix molecules are shown in red and green curves, respectively. (b) The effective spin selectivity $S(\alpha)$ calculated by the definition as in the experiment. Data with L-α-helix molecules, D-α-helix molecules, and no molecules are shown in red, green, and blue curves, respectively. The parameters used here are the same as those in Figure 2.

To evaluate the modulation strength of chiral molecules to ISHE, in Figure 3a, we show the difference between $-V_H(\alpha,+B)$ and $V_H(\alpha,-B)$, which is the summation of the Hall voltages of opposite magnetic fields $\Sigma V_H = V_H(\alpha,-B) - [-V_H(\alpha,+B)] = V_H(\alpha,+B) + V_H(\alpha,-B)$. When $\alpha = 0$ and $\pm\pi$, the spin angular momentum of the spin current is perpendicular to the helix axis of chiral molecules, so CISS cannot select spin current, and $\Sigma V_H \approx 0$ is observed for both L-α-helix and D-α-helix molecules. Since ISHE arises from the cross product of the spin current direction and spin angular momentum, Hall signals vanish at $\alpha = \pm\pi/2$ where they are parallel. Thus, we also have $\Sigma V_H \approx 0$ at $\alpha = \pm\pi/2$, even if the modulation of chiral molecules to $I_s$ reaches the strongest (see Figures 2b, 2e, and polar plot Figure S5). When $\alpha \neq 0, \pm\pi/2$, and $\pm\pi$, CISS can take effect while the Hall signal also emerges due to the cross-product relation of ISHE, resulting in $\Sigma V_H(\alpha)$ forming a shape of $\sin 2\alpha$ $(-\sin 2\alpha)$ for D-α-helix (L-α-helix) molecules. Therefore, the position with the most obvious asymmetry of Hall voltages appears at $\alpha \approx \pm\pi/4, \pm 3\pi/4$, as shown in Figure 3a, which gives a clear understanding of the experimental observations.[28]

To quantify the modulation strength of chiral molecules on the Hall voltage, we adopt the effective spin selectivity $S(\alpha) = [|V_H(\alpha, +B)| - |V_H(\alpha, -B)|]/[|V_H(\alpha, +B)| + |V_H(\alpha, -B)|]$ from Ref. [28], analogous to the spin polarization defined in prior studies.[2] As shown in Figure 3b, the behavior of $S(\alpha)$ closely resembles a sine curve, with pronounced peaks and troughs occurring at $\alpha = \pm\pi/2$. Notably, at these angles, the Hall voltages vanish, causing $S(\alpha)$ to become indeterminate (0/0), thereby limiting the availability of data for both numerical simulations and experimental measurements.[28] Except for the data near $\alpha = \pm\pi/2$, $S(\alpha)$ remains well-defined and measurable for all other $\alpha$ values. The underlying mechanism for the behavior $S(\alpha)$ is rooted in the spin angular momentum direction of the $I_s$ with respect to the chiral molecules. When $\boldsymbol{B}$ is perpendicular to the $z$-direction ($\alpha = 0, \pm\pi$), the modulation strength vanishes, as CISS cannot distinguish spins perpendicular to the molecular helix axis. Conversely, as $\alpha$ gradually approaches $\pm\pi/2$, the amplitude of $S(\alpha)$ increases, since CISS has optimal selectivity for spins aligned parallel or antiparallel to the molecular helix axis. When $0 < \alpha < \pi$ $(-\pi < \alpha < 0)$, $S(\alpha) > 0$ $(<$

0) because the L-α-helix molecules enhance the spin current in the $+z$ direction and weaken that in the $-z$-direction. The L-α-helix and D-α-helix molecules exhibit opposite modulation strength, arising from the opposite spin selectivity of different chirality in CISS. In sharp contrast, the absence of chiral molecules results in strictly zero spin selectivity $S(\alpha)$ across all $\alpha$ values, as shown in Figure 3b. The remarkable similarity between the experimentally measured spin selectivity[28] and the shape of $S(\alpha)$ depicted in Figures S2c, S2f shows that the asymmetric spin-to-charge conversion originates from the CISS of the chiral molecules.

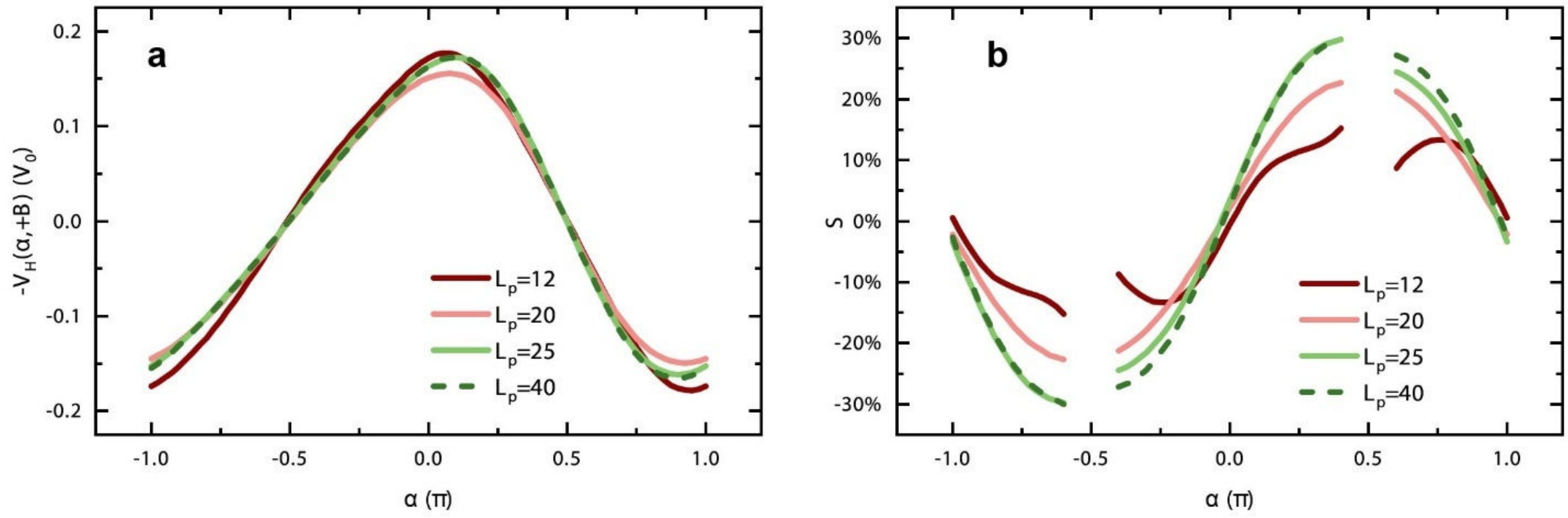

**Figure 4** The length dependence of chiral modulation with L-α-helix molecules. The minus Hall voltages $-V_H(\alpha, +B)$ [in (a)] and the effective spin selectivity $S(\alpha)$ [in (b)] with four different lengths of molecules $L_p = 12$, 20, 25, and 40 are shown in different colors. Here, we use the L-α-helix molecules as an example. Parameters used here are provided in Section S1 of Supporting Information.

In Figure 3b, the maximum of $S(\alpha)$ is about 23% with the molecule length $L_p = 20$ (the number of sites). The CISS effect becomes stronger for longer chiral molecules.[10,14,21,39] Therefore, we also compare the modulation with different $L_p$, the results are shown in Figure 4. Hall voltages with different lengths of L-α-helix molecules show similar asymmetry in Figure 4a, although their amplitudes show little difference. The modulation ability of molecules with different lengths on the ISHE is demonstrated in Figure 4b. With the increase of $L_p$, the amplitude of the sine-like curve of $S(\alpha)$ indeed increases. For example, the maximum of $S(\alpha)$ is more than 30% when the molecule length $L_p = 40$. The results for D-α-helix molecules give the opposite modulation and the same behavior of length dependence, as shown in Figure S6. In addition to the length, we also investigated the influence of dephasing strength on the regulation of ISHE by chiral molecules, which is regarded by many studies as an important part of CISS.[10,14,40,41] As shown in Figure S7, as dephasing increases from almost zero to very strong, the spin selectivity first strengthens and then weakens, which is consistent with theoretical[10,14,42] and experimental studies.[43]

When the chiral molecules self-assemble on the gold surface, their helix axis forms a tilt angle of approximately 60° with respect to the sample surface to achieve stable self-assembly.[28,44] In the above calculations, we have set the tilt angle 60° and azimuthal angle $\varphi = 0$ of the molecular helix axis (i.e., the helix axis is fixed at the $xz$-plane) as shown in Figure S1. Experiments show that the molecular helix axis also has freely valued azimuthal angles. Molecules with the same azimuthal angle form domains, and the experimental results are the combined outcomes of molecules in many different domains.[28] So below we fix the tilt angle of 60° of the helix axis and the azimuthal angles are distributed isotropically and randomly in the interval $[0,2\pi)$ as in the experiment,[28] (see Figure S8a for the azimuthal angle $\varphi$). We calculated 1000 sets of random configurations and the average results are shown in Figure S8. The asymmetric modulation of chiral molecules on the Hall voltages $V_H$ well holds. The summation of Hall voltages $\Sigma V_H = V_H(\alpha, +B) + V_H(\alpha, -B)$ under opposite $\boldsymbol{B}$ still resembles a sine curve, except for a small phase shift which attributed to the mismatch between the ISHE model and the chiral molecular model.

In this work, we present a straightforward physical interpretation of how chiral molecules modulate the spin currents and, consequently, the ISHE signal observed in the experiment.[28] Our simulations show that when the spin direction aligns closely with the CISS preferred direction, the spin current is enhanced, resulting in a stronger Hall voltage. In contrast, when the spin direction is far away from this preferred direction, both the spin current and Hall voltage are obviously reduced. This directional dependence is consistently reflected in the simulated spin currents and Hall voltages. In addition, our numerical results, including the Hall voltages of bare gold, chiral molecules, and control groups, as well as the angular-resolved spin selectivity, show excellent agreement with the experimental observations.[28] Together with the experiment, this Letter reveals a new pathway for tuning spin transport and spin-to-charge conversion via molecular chirality, offering fresh opportunities for molecular control in spintronic applications.

SUPPORTING INFORMATION

Section S1: Model details; Section S2: Sample geometry; Section S3: Details of control groups; Section S4: Supplenmentary results; Section S5: Supplementary references; Section S6: Supplementary figures (Figures S1-S8). (PDF)

AUTHOR INFORMATION

**Corresponding Author**

*Qing-Feng Sun* - International Center for Quantum Materials, School of Physics, Peking University, Beijing, 100871, China; Hefei National Laboratory, Hefei 230088, China. Email: sunqf@pku.edu.cn

**Notes**

The authors declare no competing financial interest.

ACKNOWLEDGMENT

P. L. thanks Zi-Jun Zhao for beautifying the schematics and Yue Mao for helpful discussions. This work was financially supported by National Natural Science Foundation of China (Grants No. 12374034, No. 11921005, and No. 12274466), Innovation Program for Quantum Science and Technology (2021ZD0302403), Strategic priority Research Program of Chinese Academy of Sciences (Grant No. XDB28000000), and Hunan Provincial Science Fund for Distinguished Young Scholars (Grant No. 2023JJ10058). The authors also acknowledge the High-performance Computing Platform of Peking University for providing computational resources.

# SUPPORTING INFORMATION

# Spin-to-charge conversion modulated by chiral molecules

*Peng-Yi Liu,*[1] *Tian-Yi Zhang,*[1] *Ai-Min Guo,*[2] *Yossi Paltiel,*[3,4] *Qing-Feng Sun*[1,5*]

[1]International Center for Quantum Materials, School of Physics, Peking University, Beijing, 100871, China.

[2]Hunan Key Laboratory for Super-microstructure and Ultrafast Process, School of Physics, Central South University, Changsha 410083, China.

[3]Institute of Applied Physics, The Hebrew University of Jerusalem, Jerusalem 9190401, Israel.

[4]Center for Nanoscience and Nanotechnology, The Hebrew University of Jerusalem, Jerusalem 9190401, Israel.

[5]Hefei National Laboratory, Hefei 230088, China.

*Corresponding author: sunqf@pku.edu.cn

## Section S1: Model details

### Theoretical model framework

As the experimental setup,[1] we construct a model consisting of a layer of metal with ISHE, a ferromagnetic substrate, two voltage electrodes, and self-assembled chiral molecules, as illustrated in Figures 1a and 1b. To study the electron transport, we discretize the entire system, as shown in Figure S1. Below, we introduce each component in detail, presenting the corresponding Hamiltonians and the couplings between them.

### Model of inverse spin Hall effect

The gold layer is adopted to achieve the inverse spin Hall effect (ISHE) in Ref. 1 and many other studies.[2] However, there is evidence that the spin Hall effect (SHE) of gold is mainly contributed by skew scattering,[3] which is difficult to describe exactly by simple models. Here, we adopt a SHE model similar to the Bernevig-Hughes-Zhang model,[4] to simulate the SHE and ISHE in the gold layer. In the tight-binding representation, the Hamiltonian is given by

$$H_{SHE} = \sum_{x,z} c_{x,z}^{\dagger} H_0^{SHE} c_{x,z} + \sum_{x,z} c_{x+1,z}^{\dagger} H_x^{SHE} c_{x,z} + \sum_{x,z} c_{x,z+1}^{\dagger} H_z^{SHE} c_{x,z} + h.c. \quad \text{(S1)}$$

Here, "$h.c.$" represents the Hermitian conjugation. The $y$-direction can be considered homogeneous, therefore the system is reduced to two dimensions, as shown in Figure S1. $c_{x,z} = [c_{x,z;\mathscr{s},\uparrow}, c_{x,z;\mathscr{p},\uparrow}, c_{x,z;\mathscr{s},\downarrow}, c_{x,z;\mathscr{p},\downarrow}]^T$, with $T$ denoting transpose. $(x, z)$ is the label of a site in the gold layer, and $c_{x,z;\mathscr{s}(\mathscr{p}),\uparrow(\downarrow)}$ is the annihilation operator of the electron at $(x, z)$ with pseudospin $\mathscr{s}$ or $\mathscr{p}$ and spin parallel or antiparallel to the direction of $\boldsymbol{B}$. $c_{x,z;\mathscr{s}(\mathscr{p}),\uparrow(\downarrow)}^{\dagger}$ is the corresponding creation operator. $H_0^{SHE}$, $H_x^{SHE}$ and $H_y^{SHE}$ are given by,

$$\begin{aligned} H_0^{SHE} &= E_0(s_0 \otimes \sigma_0) - M(s_\alpha \otimes \sigma_0) - (m - 4t)(s_0 \otimes \sigma_y) \\ H_x^{SHE} &= -t(s_0 \otimes \sigma_y) + \frac{it}{2}(s_0 \otimes \sigma_x) \\ H_z^{SHE} &= -t(s_0 \otimes \sigma_y) + \frac{it}{2}(s_y \otimes \sigma_z). \end{aligned} \quad \text{(S2)}$$

$E_0$ is the on-site energy of the gold. $M$ is a very small Zeeman term due to the magnetic field. In the spin space, $s_0$, $s_x$, $s_y$, and $s_z$ represent the unit matrix, the Pauli matrix in the $x$, $y$, and $z$ directions, respectively. $s_\alpha$ is the Pauli matrix with spin towards the direction of $\boldsymbol{B}$. $\sigma_0$, $\sigma_x$, $\sigma_y$, and $\sigma_z$ represent Pauli matrices in the pseudospin space, which comes from the orbital degrees of freedom of electrons in atoms of the ISHE system. The mass term $m$ and the hopping energy $t$ are the parameters in the Bernevig-Hughes-Zhang model.

The ferromagnetic substrate (also seen as an electrode) and two electrodes are described by a simpler Hamiltonian $H_E = \sum_{x',z'} a_{x',z'}^{\dagger} H_0^E a_{x',z'} + \sum_{x',z'} a_{x'+1,z'}^{\dagger} H_x^E a_{x',z'} + \sum_{x',z'} a_{x',z'+1}^{\dagger} H_z^E a_{x',z'} + h.c.$, with $H_0^E = E_0(s_0 \otimes \sigma_0) - M(s_\alpha \otimes \sigma_0)$ and $H_x^E = H_z^E = -t(s_0 \otimes \sigma_0)$. $(x', z')$ is the label of a site in electrodes. $a_{x',z'}^{\dagger}$ $(a_{x',z'})$ represents the creation (annihilation) operator of electrons in the electrodes at the site $(x', z')$ and has the same basis as $c_{x,z}^{\dagger}$ $(c_{x,z})$. $t$, $E_0$, and $M$ are the same as $H_{SHE}$. The coupling Hamiltonian between the gold layer and electrodes is $H_C^E = \sum'_{x,z;x',z'} t_c^E c_{x,z}^{\dagger}(s_0 \otimes \sigma_0)\, a_{x',z'} + h.c.$, where the summation $\Sigma'_{x,z;x',z'}$ only operates when $(x, z)$ and $(x', z')$ are the nearest neighbors (blue bonds in Figure S1). The gold layer in our model has $N = 20$ sites in the $x$-direction and $L = 10$ sites in the $z$-direction. The detailed size of our model is shown in Figure S1.

## Chiral molecule

The chiral molecules we used are the $\alpha$-helical proteins, which can be described by a tight-binding Hamiltonian including SOC,[5–7] $H_M = H_{mol} + H_{SOC}$.

$$
\begin{aligned}
H_{mol} &= \sum_{n=1}^{N_p}\left[\sum_{l=1}^{L_p} d_{n,l}^{+}\left(E_p s_0 - M s_\alpha\right) d_{n,l} + \sum_{l=1}^{L_p-1}\sum_{j=1}^{L_p-l} t_j d_{n,l}^{+} s_0 d_{n,l+j} + h.c.\right], \\
H_{SOC} &= \sum_{n=1}^{N_p}\sum_{l=1}^{L_p-1}\sum_{j=1}^{L_p-l} 2\mathrm{i} t_{sj} \cos\varphi_{lj}^{-}\, d_{n,l}^{+} s_{lj} d_{n,l+j} + h.c.
\end{aligned}
\tag{S3}
$$

$d_{n,l}^{+} = [d_{n,l,\uparrow}^{+}, d_{n,l,\downarrow}^{+}]$ is the creation of electrons at site $l$ of the $n$-th molecule. Each molecule has $L_p$ sites (i.e., the molecule length is $L_p$), and there are $N_p = N/2$ molecules coupled to the surface of the gold. $M$ is the Zeeman term. $E_p$ is the on-site energy of molecules. $t_j = t_1 e^{-(l_j - l_1)/l_c}$ is the long-range hopping energy[5,8] between site $l$ and site $l + j$. Similarly, $t_{sj} = t_{s1} e^{-(l_j - l_1)/l_c}$ is the corresponding SOC. $t_j$ and $t_{sj}$ decays with the exponent $l_c$. The geometry of the $\alpha$-helical molecule can be described by the radius $R$, the twist angle between the nearest sites $\Delta\varphi$, and the pitch of the helix $\Delta h$. Then, the Euclidean distance between sites $l$ and $l + j$ is $l_j = \sqrt{[2R\sin(j\Delta\varphi/2)]^2 + (j\Delta h)^2}$, and the corresponding space angle $\theta_j = \arccos[2R\sin(j\Delta\varphi/2)\, /\, l_j]$. According to our previous result,[5,9] $s_{lj} = (s_x' \sin\varphi_{lj}^{+} - s_y' \cos\varphi_{lj}^{+})\sin\theta_j + s_z' \cos\theta_j$, with $\varphi_{lj}^{\pm} = [(n + j) \pm n]\Delta\varphi/2$. This model can successfully simulate the chiral-induced spin selectivity (CISS) effect, which we attribute to the interference of multi-path wave functions (see Ref. 10). Considering that experiments have shown that the helical molecules form a tilt angle of about $\pi/6$ with the $z$-direction when self-assembled on the surface of gold,[1,11] we rotate the Pauli matrix of the helical molecules accordingly $s'_{x(y,z)} = \exp(+\mathrm{i}\pi s_y/12)\, s_{x(y,z)} \exp(-\mathrm{i}\pi s_y/12)$. That is, the molecules are set at a tilt angle of $\pi/6$ with the $z$-direction and an azimuthal angle $\varphi = 0$ (i.e., the helical axes of molecules are fixed at the $xz$-plane) as shown in Figure S1. In the experiment, the azimuthal angles of chiral molecules in the $xy$-plane are considered to be isotropic distributed rather than fixed.[1] After considering the azimuthal freedom, our model is still valid, as we discussed in Section S4 and Figure S8.

The coupling Hamiltonian between the gold layer and molecules is $H_C^M$.

$$
H_C^M = \sum_{n=1}^{N_p} t_c^M d_{n,1}^{+}\, [s_0 \otimes (1, 1)]\, \left[c_{(2n-1,L)} + c_{(2n,L)}\right] + h.c. \tag{S4}
$$

Where $z = L$ represents the surface of the gold. The $n$-th molecule is coupled to the $(2n-1)$- and $2n$-th site of the surface, as shown in Figure S1.

**Dephasing effect**

In real systems, impurities with degrees of freedom, electron-electron interaction, and electron-phonon interaction inevitably cause the loss of phase memory of electrons.[12] Similarly, it is generally believed that dephasing plays a role in molecules as well.[13,14] Consequently, we incorporate the dephasing effect into both the ISHE system and the chiral molecules. This is achieved by utilizing the widely adopted Büttiker virtual electrodes, which introduce a phase-breaking mechanism to electrons.[15–17] We attach a virtual electrode to each site of the gold layer and molecules. Electrons can go into and out of the virtual electrodes with the loss of phase memory. The Hamiltonian of the virtual electrodes is described by $H_D = \sum_{x,z}\sum_k \varepsilon(k)\Psi^\dagger_{(x,z)k}\Psi_{(x,z)k} + \sum_{n,l}\sum_k \varepsilon(k)\Psi^\dagger_{(n,l)k}\Psi_{(n,l)k}$, where $\Psi^\dagger_{(x,z)k}$ is the creation operator of electrons in the virtual electrode coupled to the gold site $(x,z)$ with momentum $k$. $\Psi^\dagger_{(n,l)k}$ is the creation operator in the virtual electrode coupled to the $l$-th site of the $n$-th molecule with momentum $k$. $\Psi^\dagger_{(x,z)k} = (\Psi^\dagger_{(x,z)k,s\uparrow}, \Psi^\dagger_{(x,z)k,p\uparrow}, \Psi^\dagger_{(x,z)k,s\downarrow}, \Psi^\dagger_{(x,z)k,p\downarrow})$, and $\Psi^\dagger_{(n,l)k} = (\Psi^\dagger_{(n,l)k,\uparrow}, \Psi^\dagger_{(n,l)k,\downarrow})$. $\varepsilon(k)$ is the spectrum of each virtual electrode. The coupling between virtual electrodes and real sites in the system is $H_C^D$.

$$H_C^D = \sum_k \sum_{(x,z)} t_d \Psi^\dagger_{(x,z)k}(s_0 \otimes \sigma_0)c_{x,z} + \sum_k \sum_{n,l} t_d \Psi^\dagger_{(n,l)k} s_0 d_{n,l} + h.c. \qquad \text{(S5)}$$

**Transport formulation**

According to the three subsections above, the total Hamiltonian of our system includes the electrodes, SHE system, chiral molecules, virtual electrodes, and their coupling $H_{tot} = H_E + H_{SHE} + H_M + H_D + H_C^E + H_C^M + H_C^D$. We label all electrodes with integers $p = 1,2,3, ....$ As shown in Figure S1, we set $p = 1$ for the ferromagnetic substrate, $p = 2, 3$ for electrodes measuring the Hall voltage, and $p > 3$ for virtual electrodes.

The calculation of electron transport is given by the Landauer-Büttiker formula,[18]

$$I_p^s = \frac{e^2}{h}\sum_{q,s'} T_{pq}^{ss'}\left(V_p^s - V_q^{s'}\right) \qquad \text{(S6)}$$

where $I_p^s$ is the charge current with spin-$s$ from the $p$-th electrode to the system. $s, s' = \uparrow, \downarrow$ represents the spin parallel or anti-parallel with $\boldsymbol{B}$. $V_p^s$ ($V_q^{s'}$) represents the voltage of the $p$- ($q$-) th electrodes with spin-$s$ ($s'$). $T_{pq}^{ss'}$ is the transmission coefficient of electrons from the $q$-th electrode with spin-$s'$ to the $p$-th electrode with spin-$s$. With the non-equilibrium Green function, the transmission coefficient is given by $T_{pq}^{ss'} = \mathrm{Tr}[\boldsymbol{\Gamma}_\mathrm{p}^\mathrm{s}(E_F)\mathbf{G}^r(E_F)\boldsymbol{\Gamma}_\mathrm{q}^{\mathrm{s}'}(E_F)\mathbf{G}^a(E_F)]$.[19] The retarded Green function $\mathbf{G}^r(E) = [\mathbf{G}^a(E)]^\dagger = [E - \mathbf{H}_{SHE} - \mathbf{H}_M - \Sigma_p \boldsymbol{\Sigma}_{p,s}^r(E)]^{-1}$, and the linewidth function $\boldsymbol{\Gamma}_\mathrm{p}^\mathrm{s}(E) = \mathrm{i}[\boldsymbol{\Sigma}_{p,s}^r(E) - \boldsymbol{\Sigma}_{p,s}^a(E)]$, where the bold letters represent the matrix under the tight-binding representation. The Fermi energy is set to be zero $E_F = 0$. $\mathbf{\Sigma}^r_{p,s}(E) = [\mathbf{\Sigma}^a_{p,s}(E)]^\dagger$ is the retarded self-energy introduced by the coupling of the $p$-th electrodes with spin-$s$. For the first three electrodes, we use a numerical method to obtain their self-energy (*53*). For virtual electrodes, $\mathbf{\Sigma}^r_{p>3,s} = -\mathrm{i}\pi\rho t_d^2 = -\mathrm{i}\Gamma_\mathrm{d}/2$ is energy independent, with $\rho$ being the density of states in each virtual electrode and $\Gamma_\mathrm{d}$ the dephasing strength.[16,17]

Considering the Fermi energy splitting of FMR, a spin bias is applied to the first electrode $V_1^\downarrow = -V_1^\uparrow = V_0$. While for other electrodes, $V_{p>1}^\uparrow = V_{p>1}^\downarrow$. Since there is no net charge current for voltage electrodes (the 2nd and 3rd electrodes) and the dephasing process, $I_{p>1}^\uparrow + I_{p>1}^\downarrow = 0$. With these conditions, $I_p^s$ and $V_p^s$ can be solved from Eq. (6) for any $p$ and $s$. The spin current through the ferromagnetic/gold interface is $I_s = -\hbar(I_1^\uparrow - I_1^\downarrow)/(2e)$, and the Hall voltage is given by $V_H = V_2^\uparrow - V_3^\uparrow$. The charge current through the ferromagnetic/gold interface is $I_c = I_1^\uparrow + I_1^\downarrow$, which is zero exactly due to the current conservation.

### Model parameters

Unlike the original application of the Bernevig-Hughes-Zhang model in quantum SHE (*33*), we set $m = 0.2$, $t = 1$, and $E_0 = -0.61$ to ensure that the Fermi surface lies in the conduction band, resulting in a metallic SHE.[20] The geometric parameters of chiral molecules we used are the same as our previous work:[5] $R = 0.25$nm, $\Delta\varphi = 5\pi/9$, $\Delta h = 0.15$nm, and $l_c = 0.09$nm. Considering that the hopping energy in gold is approximately on the order of electron volts, while the hopping in organic molecules is on the order of 100meV,[21] we set $t_1 = t/25 = 0.04t$. We use a smaller SOC than our previous work $t_{s1} = 0.05t_1$, which is about a few meV and close to the experimental values.[22,23] The dephasing strength is set to be energy-independent $\Gamma_\mathrm{d} = 0.02t_1$, and we explore the dephasing dependence of the effect in Figure S7. The small Zeeman term is $M = 0.03t_1 = 0.0012t$ for $H_{SHE}$, $H_E$, and $H_M$. The coupling strengths are given by $t_c^E = 0.5t$ and $t_c^M = 0.64t$. In Figures 2 and 3, we use $L_p = 20$ and $E_p = 0.57t_1$. In Figure 4, other lengths of molecules and the on-site energy of molecules are $(L_p, E_p) = (12, 0.36t_1)$, $(25, 0.73t_1)$, and $(40, 1.15t_1)$.

## Section S2: Sample geometry

### Geometric details

The microscopic schematic diagram of our model is shown in Figure S1. In real calculations, we use the system size $N = 20$ and $L = 10$ instead of the size in the schematic diagram. The gold layer in the real experiment[1] is about 4nm ~ 10 lattice constants of gold, which is the same as $L$. Each chiral molecule maintains a 30-degree angle with the $z$-direction to self-assemble stably, as described in the main text. As a reasonable simplification, we adopt a square lattice and couple the first site of each molecule to the metal surface (see the Methods in the main text). The $y$- direction of the lattice is considered homogeneously and does not appear directly in the calculation.

**Symmetry analysis without chiral molecules**

When no molecule is involved, the Hamiltonian of the system is given by $H' = H_{SHE} + H_E + H_C^E + H_D' + H_C^D{}'$, where $H_D' + H_C^D{}'$ contains only virtual electrodes coupled to the gold layer (the ISHE system). Then, we can define four types of symmetry operations. (I) The unitary rotation operators in spin space $\hat{\mathcal{U}}(s_x, \theta) = \exp(-\mathrm{i}\theta s_x/2)$. (II) The unitary rotation operators in pseudospin space $\hat{\mathcal{U}}(\sigma_y, \theta) = \exp(-\mathrm{i}\theta\sigma_y/2)$. (III) The mirror operation about the red dashed line $\hat{\mathcal{M}}$ in Figure S1. (IV) The operation $\hat{\mathcal{P}}$ replaces the Zeeman term $M$ with $-M$. It can be verified that all five terms in $H'$ remain invariant under the joint operation $\hat{\mathcal{W}} = \hat{\mathcal{U}}(s_x, \pi) \circ \hat{\mathcal{U}}(\sigma_y, \pi) \circ \hat{\mathcal{M}} \circ \hat{\mathcal{P}}$. After applying $\hat{\mathcal{W}}$ to the system, the spin direction of the spin current reverses and the positions of the two electrodes for measuring Hall voltage are interchanged. Therefore, $-V_H(\alpha, +B) = V_H(\alpha, -B)$ is strictly true without the participation of molecules.

## Section S3: Details of control groups

As mentioned in the main text, we simulate the racemic mixture by assigning each molecule a 50% probability of being L-α-helix and a 50% probability of being D-α-helix. The calculation results are averaged over 1000 sets of random configurations, as shown in Figures 2h and 2i. Here, we present the standard deviation of 1000 sets of data to support the results in the main text, as the error bars in Figure S3. It can be observed that the data for positive and negative magnetic fields slightly deviate beyond the coverage range of the error bars. However, the magnitude of deviations remains relatively small, which suggests the racemic mixture cannot introduce chiral-induced asymmetry. These minor deviations are attributed to the mismatch in pseudospin (orbital degree of freedom) between the molecular model and the ISHE model, which breaks the pseudospin rotation symmetry.

Another control group uses achiral molecules, whose Hamiltonian only includes simple one-dimensional chains. $H_M = \sum_{n=1}^{N_p} \left[ \sum_{l=1}^{L_p} d_{n,l}^+ (E_p s_0 - M s_\alpha) d_{n,l} + \sum_{l=1}^{L_p - 1} t_1 d_{n,l}^+ s_0 d_{n,l+1} + h.c. \right]$. The symbol definitions here are the same as those in the main text. We use $t_1 = 1/25$, $E_p = 0.2t_1$, and $M = 0.03t_1$ for Figures 2k, 2l, S4b.

We can also use the effective spin selectivity $S(\alpha)$ defined in the main text to test the control groups. The results are shown in Figures S4a and S4b, with parameters the same as those in Figure 2. Compared with the notable spin selectivity of chiral molecules in Figure 3b, the spin selectivity of the two control groups is close to 0, which can support the asymmetry in Figure 2a-2f is caused by chirality. The slight deviation of $S(\alpha)$ from 0 arises from the mismatch between $H_{SHE}$ containing pseudospin and $H_M$ without pseudospin. If we use molecules containing pseudospin, the control group can also maintain $S(\alpha) = 0$ exactly. For example, if $H_M = \sum_{n=1}^{N_p} \left[ \sum_{l=1}^{L_p} d_{n,l}^+ (E_p s_0 \otimes \sigma_0 - M s_\alpha \otimes \sigma_0) d_{n,l} + \sum_{l=1}^{L_p - 1} t_1 d_{n,l}^+ (s_0 \otimes \sigma_0) d_{n,l+1} + h.c. \right]$ and $H_C^M = \sum_{n=1}^{N_p} t_C^M d_{n,1}^+ (s_0 \otimes \sigma_0) \left[ c_{(2n-1,L)} + c_{(2n,L)} \right] + h.c.$, the spin selectivity will become zero with any $t_C^M$ and $E_p$, as shown in Figure S4c (here, we use the same $t_C^M$ and $E_p$ as chiral molecules in Figure 2).

## Section S4: Supplenmentary results

**The Hall voltages and spin currents in the polar coordinate systems**

Plotting the absolute value of Hall voltages $|V_H(\alpha, +B)|$ in a polar coordinate system can provide a clear comparison of the different modulation effects of different chirality on the ISHE. As shown in Figure S5a, $|V_H(\alpha, +B)|$ with the racemic mixture shows no difference for the $\pm z$-direction. But the results of L-α-helix (D-α-helix) are notably shifted to $\alpha = \pi/2$ $(-\pi/2)$. Similar shifts can be observed from the results of spin current $I_s(\alpha, +B)$, as shown in Figure S5b. These asymmetries are due to the fact of CISS that L-α-helix molecules we used prefer spin currents with spin towards the $+z$-direction, while D-α-helix molecules prefer spin currents with spin towards the $-z$-direction. As we discussed in the main text: The asymmetry of $|V_H(\alpha, +B)|$ is most obvious at $\alpha \approx \pm\pi/4, \pm3\pi/4$ and disappears at $\alpha = 0, \pm\pi/2, \pm\pi$. The asymmetry of $I_s(\alpha, +B)$ is most obvious at $\alpha = \pm\pi/2$ and disappears at $\alpha = 0, \pm\pi$.

**The length dependence of chiral modulation with D-α-helix molecules**

In the main text, the length dependence of the modulation strength for L-α-helix molecules is discussed in Figure 4. In this Supporting Information, we also show the Hall voltages and spin selectivity of D-α-helix molecules with four different lengths, as shown in Figure S6. The Hall voltage of molecules with different lengths all show similar asymmetry (Figure S6a), and the spin selectivity is positively correlated with the molecular length (Figure S6b). Compared with the results of L-α-helix in Figure 4, the results for D-α-helix molecules give the opposite modulation and the same behavior of length dependence.

**The dephasing dependence of ISHE and spin selectivity with L-α-helix molecules**

Dephasing is considered to play an important role in CISS, so we explored the influence of dephasing strength on the effect. As shown in Figure S7a, ISHE with different dephasing strengths all show chiral-induced asymmetry, although the specific values are slightly different. Correspondingly, we calculated the spin selectivity $S$ at each dephasing strength, as shown in Figure S7b. It can be seen that the dephasing strength modulates the magnitude of this effect. As shown in Figure S7c, with the dephasing increasing from almost zero to very strong, $S$ first increases and then decreases, which aligns well with both theoretical[5,9,24] and experimental[25] studies of CISS. Within a wide range of parameters, CISS can notably modulate ISHE.

**The average of the random azimuthal angles**

When self-assembling on the gold surface, helical molecules do not align their helical axes perpendicularly to the surface but rather form an angle of approximately 60° with respect to the sample surface.[1,11] Consequently, in our numerical simulations, we similarly maintain a 30° angle between the helical axes of the chiral molecules and the $z$-axis. Specifically, considering each chiral molecule with its initial helical axis oriented towards the $+z$-direction, we rotate it by 30° around the $y$-axis to achieve the configuration depicted in Figure S1. Since SOC in chiral molecules is dependent on the orientation of the helical axis,[5] this rotation influences the Pauli matrices within $H_{SOC}$, through the transformation mentioned in the main text $s'_{x(y,z)} = \exp(+\mathrm{i}\pi s_y/12)\, s_{x(y,z)} \exp(-\mathrm{i}\pi s_y/12)$.

Experiments show that the helical axes of these molecules also have freely valued azimuthal angles. Molecules with the same azimuthal angle form domains, and the experimental results are the combined outcomes of molecules in many different domains.[1] So, we introduce an additional rotation of each chiral molecule around the $z$-axis based on the settings in Figure S1. This rotation displaces the helical axis out of the $xz$-plane, resulting in an azimuthal angle $\varphi$ in the $xy$-plane, as shown in Figure S8a. As a result, the Pauli matrices are correspondingly rotated $s''_{x(y,z)} = \exp(+\mathrm{i}\varphi s_z/2)\, s'_{x(y,z)} \exp(-\mathrm{i}\varphi s_z/2)$. Since the azimuthal angles of the helical axes are isotropically distributed in the experimental scale,[1] we calculated 1000 sets of random configurations. In each randomly generated configuration, the azimuthal angle of each chiral molecule is uniformly and randomly chosen from the interval $[0,2\pi)$. The average results of 1000 sets of configurations are shown in Figure S8.

Figures S8b and S8c respectively show the average Hall voltages of L-α-helix and D-α-helix molecules. Comparing with Figures 2c and 2f, these results show small deformation and similar asymmetries. This indicates that the model in the main text is reasonable for explaining the modulation of chiral molecules even if random azimuthal angles are considered. These asymmetries can also be clearly shown by plotting $|V_H(\alpha, +B)|$ in a polar coordinate system. Similar to the results of fixed azimuthal angle (shown in Figure S5a), the average results of random azimuthal angles (shown in Figure S8d) indeed show a shift to $\alpha = \pi/2$ $(-\pi/2)$ for L-α-helix (D-α-helix) molecules, as a result of CISS. The magnitude of the shift is slightly reduced due to the averaging of molecules whose helical axes are not in the $xz$-plane, but the shape remains very consistent with Figure S5a. At last, we compare the summation of Hall voltages $\Sigma V_H = V_H(\alpha, +B) + V_H(\alpha, -B)$ under opposite $\boldsymbol{B}$ between fixed and average random azimuthal angles, as shown in Figures S8e and Figures S8f. After the average of random azimuthal angles, both the results of L-α-helix and D-α-helix molecules show the same behavior as the results with the fixed azimuthal angle, except for slight deformation and reduced amplitude. We attribute this small deformation to the mismatch between the SHE model and the chiral molecular model, similar to the discussion in the control groups. In summary, Figure S8 shows that our calculation results and discussion in the main text can explain the experimental results, even with the variability in azimuthal angles.

## Section S5: Supplementary references

## Section S6: Supplementary figures

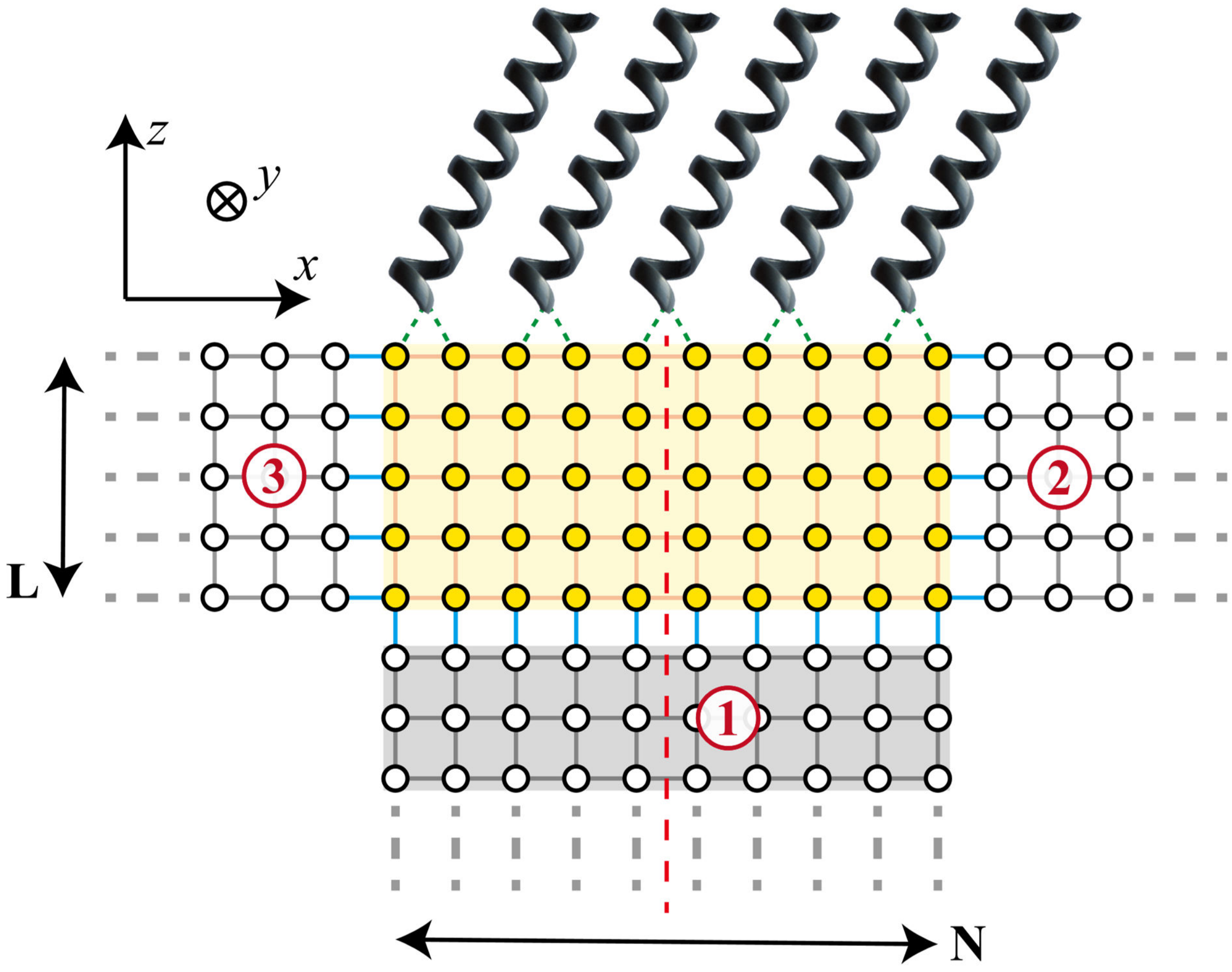

**Figure S1. The microscopic schematic diagram of the model.** The gold atoms and electrode atoms are represented by yellow and white dots, respectively. The solid orange lines, solid gray lines, solid blue lines, and dashed green lines represent the coupling within gold, within electrodes, between gold and electrodes, and between gold and molecules, respectively. The coordinate here is the same as that in Figure 1a in the main text. The electrode covered in gray represents the ferromagnetic substrate, labeled with a red "1". The two electrodes in the $x$-direction, labeled with "2" and "3", are used to measure the Hall voltage. All three electrodes extend along the dashed grey lines. The model except for the molecules is geometrically symmetric about the red dashed line. In this schematic diagram, the width of the gold layer in the $x$-direction is $N = 10$, and the width in the $z$-direction is $L = 5$. In real calculations, we fix $N = 20$ and $L = 10$.

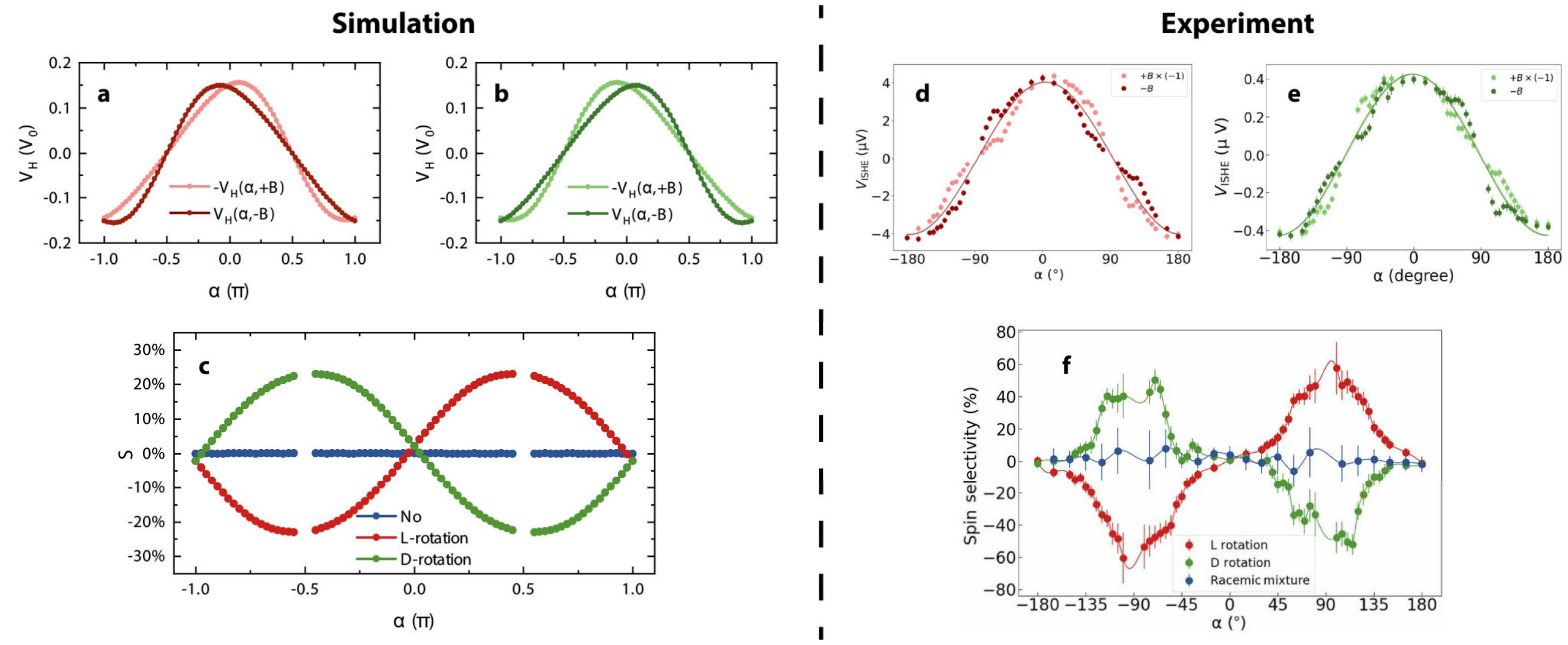

**Figure S2. The comparison of the same physical quantities in the simulation and experiment.** (a), (b), and (c) are the simulated ISHE signal with L-α-helix molecules, ISHE signal with D-α-helix molecules, and spin selectivity, respectively. (d), (e), and (f) are the corresponding experimental results. The simulation and experimental results show a high degree of similarity. (a-c) are the same as Figures 2c, 2f, and 3b in the main text. (d-f) are reproduced from Ref. 1 under the terms of the Creative Commons CC BY license.

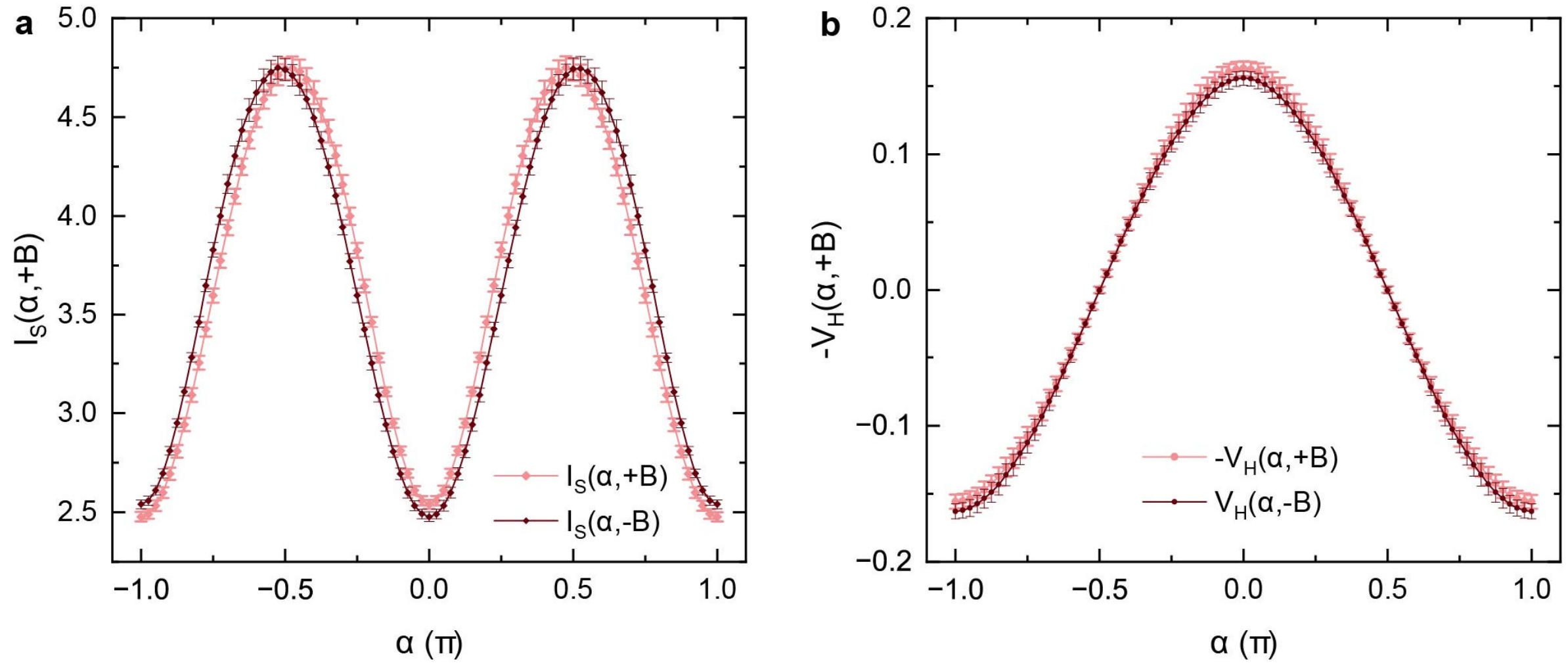

**Figure S3. The data of racemic mixture with error bar.** (a and b) are the averaged results of the spin currents and ISHE voltages, the same as Figures 2h and 2i, respectively, with error bars show the standard deviation of 1000 configurations.

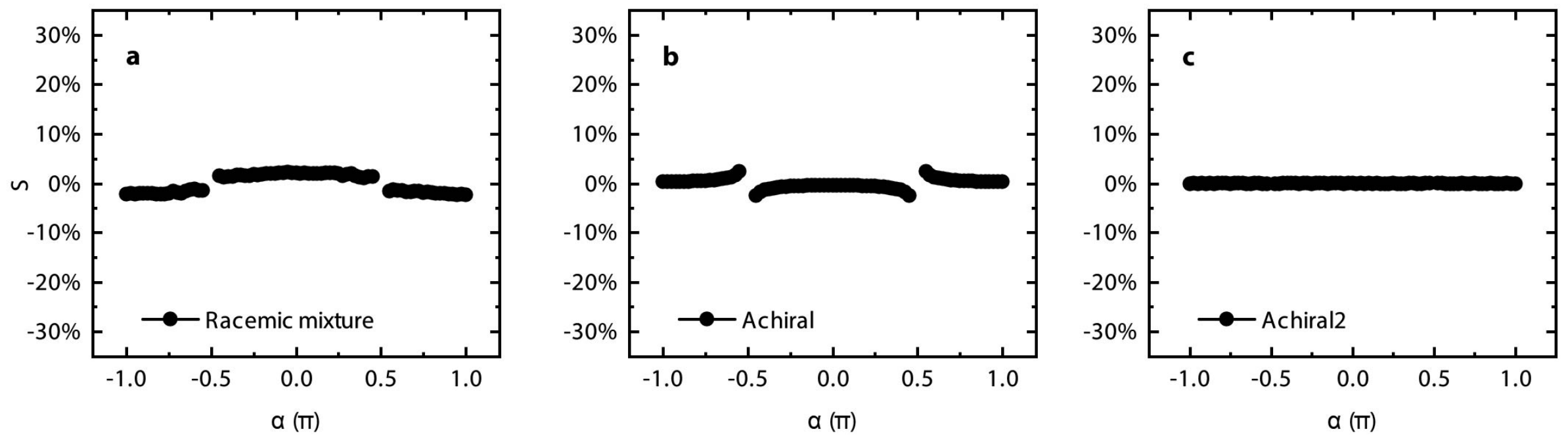

**Figure S4 The effective spin selectivity of control groups.** (a) The effective spin selectivity $S(\alpha)$ with the racemic mixture. (b) $S(\alpha)$ with the achiral molecules. (c) $S(\alpha)$ with achiral molecules which have the pseudo-spin degree of freedom. The parameters used in Figure S4 are mentioned in Section S3.

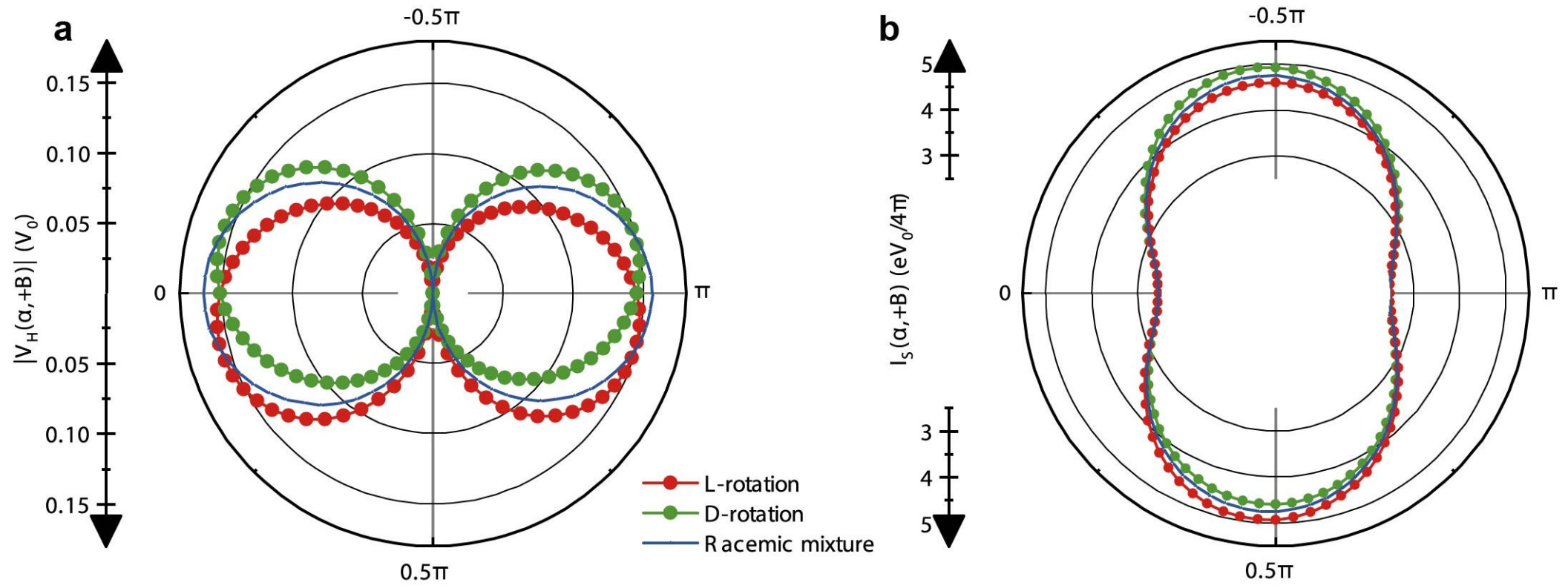

**Figure S5 The Hall voltages and spin currents in the polar coordinate systems.** (a) The absolute values of Hall voltages $|V_H(\alpha, +B)|$ with L-α-helix, D-α-helix, and racemic mixed molecules are shown in red, green, and blue curves, respectively. (b) The spin currents $I_S(\alpha, +B)$ with L-α-helix, D-α-helix, and racemic mixed molecules are shown in different colors. The results of L- (D-) α-helix molecules show a preference for the spin current and Hall effect with $\alpha = \pi/2$ $(-\pi/2)$. The results with the racemic mixture are almost the average of those with L-α-helix and D-α-helix molecules. The parameters used here are the same as in Figure 2.

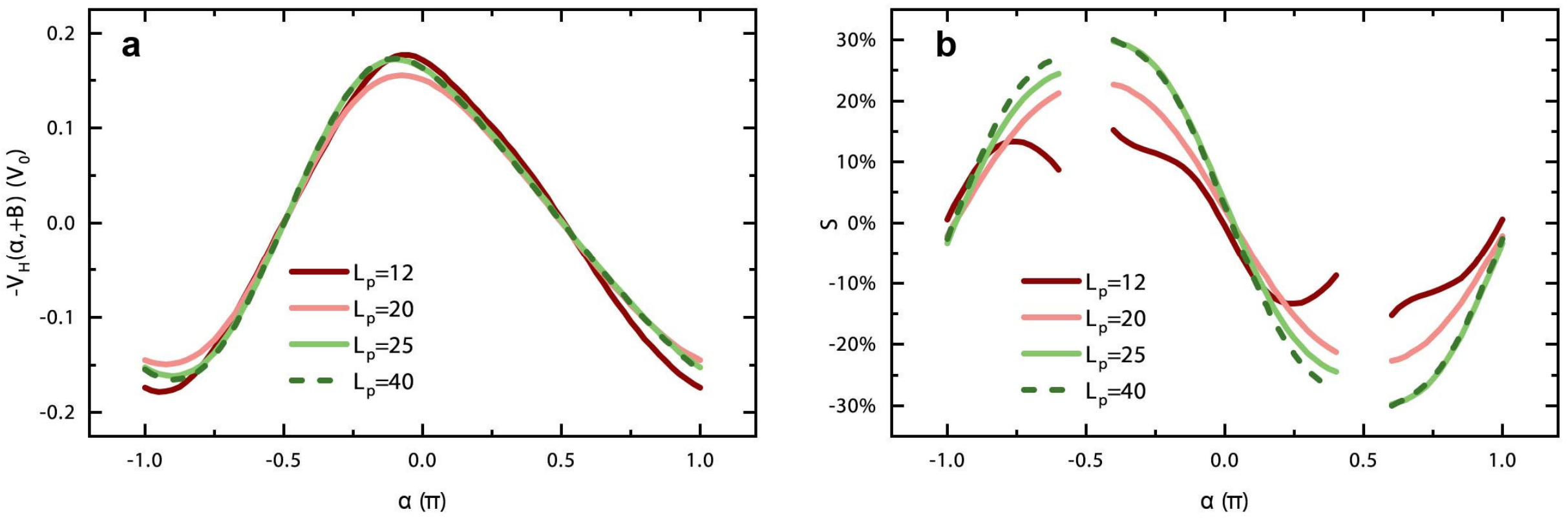

**Figure S6 The length dependence of the modulation with D-α-helix molecules.** The minus Hall voltages $-V_H(\alpha,+B)$ [in (a)] and the effective spin selectivity $S(\alpha)$ [in (b)] with four different lengths of molecules $L_p = 12$, 20, 25, and 40 are shown in different colors. Here, we use the D-α-helix molecules and the parameters are the same as Figure 4 in the main text.

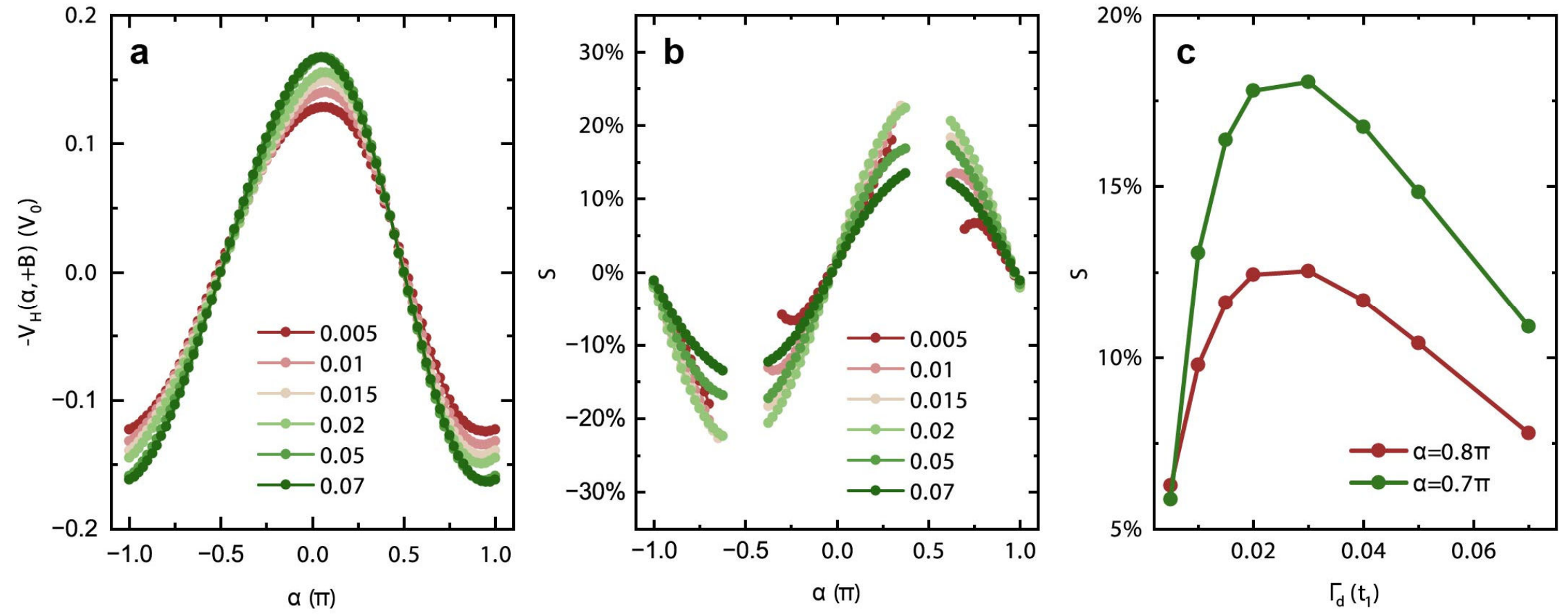

**Figure S7 The dephasing dependence of ISHE and spin selectivity with L-α-helix molecules.** The minus Hall voltages $-V_H(\alpha,+B)$ [in (a)] and the effective spin selectivity $S(\alpha)$ [in (b)] with different dephasing strength $\Gamma_d$ from 0.005 to 0.07 are shown in different colors. (c) The spin selectivity of two fixed $\alpha$ varies with the dephasing strength $\Gamma_d$. It can be seen that as dephasing increases from almost zero to very strong, the spin selectivity first strengthens and then weakens. Here, we use the L-α-helix molecules and the other parameters are the same as Figures 2 and 3 in the main text.

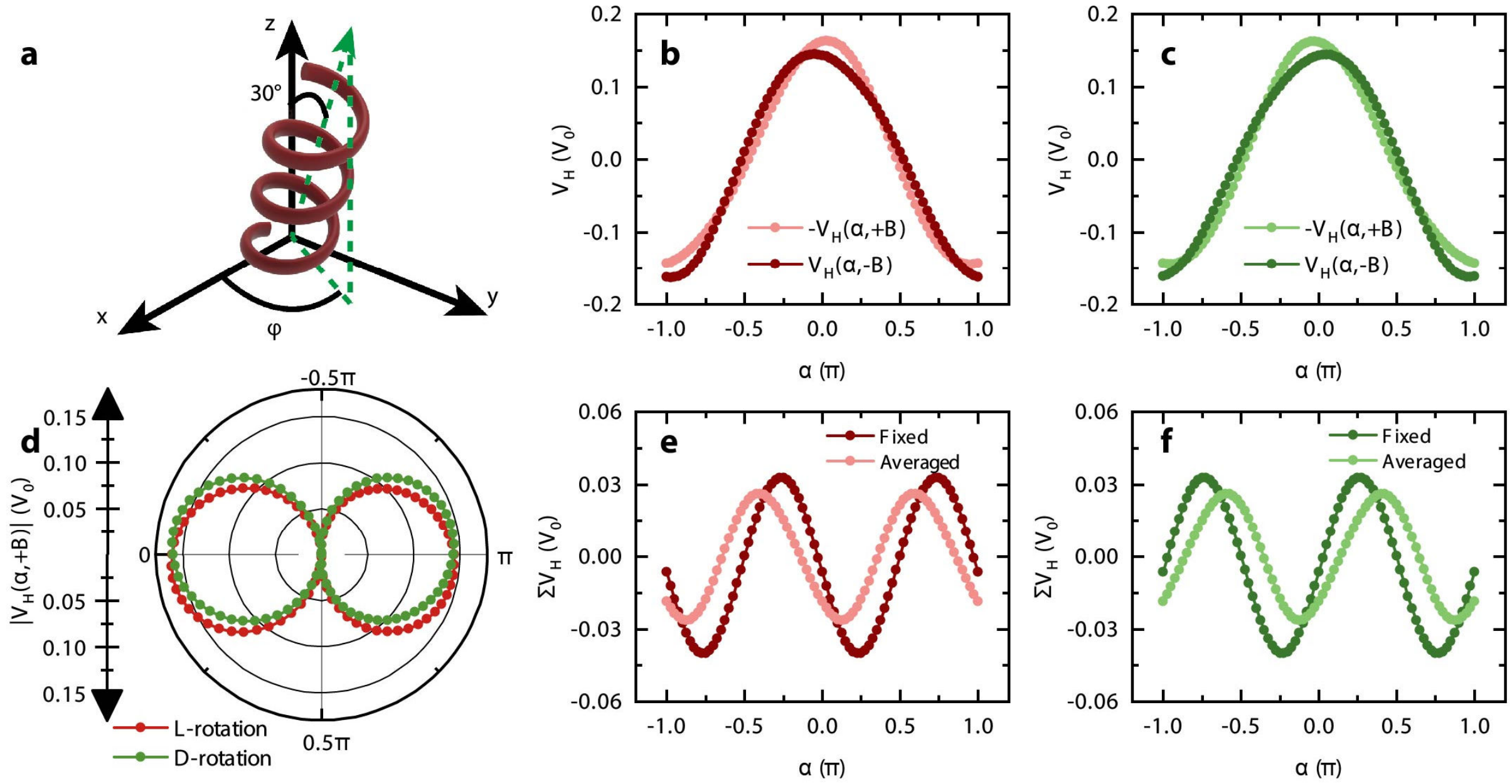

**Figure S8** The average results of chiral molecules with random azimuthal angles. (a) The schematic diagram of a self-assembled L-α-helix molecule, with the green arrow showing the helical axis. The orientation of the coordinate system here is identical to that in Figure 1 and Figure S1. The molecular helical axis maintains 30° to the $z$-axis and an azimuthal angle $\varphi$ within the $xy$-plane. In Figure S1, all molecules have $\varphi = 0$ and are fixed in the $xz$-plane. In Figure S8, each molecule has a random $\varphi \in [0,2\pi)$, and the results are averaged according to Section S4. The average Hall voltages $V_H$ (under $\pm B$) of L-α-helix and D-α-helix molecules with random azimuthal angles are shown in (b) and (c), respectively. (d) provides the averaged absolute value of the Hall voltages in a polar coordinate. Data with L-α-helix and D-α-helix molecules are shown in red and green curves, respectively. (e) and (f) show the comparison of $\Sigma V_H$ between the results of fixed azimuthal angle and the average results of random azimuthal angle. The dark red (green) curve shows the results of L-α-helix (D-α-helix) molecules with fixed azimuthal angle, and light red (green) shows the average results of L-α-helix (D-α-helix) molecules with random azimuthal angles. The parameters used here are the same as those in Figures 2a-2f, and the average details are provided in the Section S4.